\documentclass[10pt]{emulateapj}
\usepackage{apjfonts}

\newcommand{\pfrac}[2]{\left(\frac{#1}{#2}\right)}

\def\etal{{et al.~}}
\def\eps{\epsilon}

\slugcomment{ }

\shorttitle{The upstream magnetic field of GRB shocks}
\shortauthors{Li \& Waxman}

\begin{document}

\title{The upstream magnetic field of collisionless GRB shocks}

\author{Zhuo Li and Eli Waxman}
\affil{Physics Faculty, Weizmann Institute of Science, Rehovot 76100, Israel}

\begin{abstract}
Gamma-ray burst (GRB) afterglow emission is believed to be produced by synchrotron
emission of electrons accelerated to high energy by a relativistic collisionless
shock propagating into a weakly magnetized plasma. Afterglow observations have been
used to constrain the post-shock magnetic field and structure, as well as the accelerated
electron energy distribution. Here we show that X-ray afterglow observations on day
time scale constrain the
pre-shock magnetic field to satisfy $B>0.2(n/1\,{\rm cm}^{-3})^{5/8}$~mG, where $n$ is
the pre-shock density. This suggests that either the shock propagates into a highly
magnetized fast,
$v\sim10^3{\rm km/s}$, wind, or that the pre-shock magnetic field is strongly
amplified, most likely by the streaming of high energy shock accelerated particles.
More stringent constraints may be obtained by afterglow observations at high photon
energy at late, $>1$~d, times.
\end{abstract}


\section{Introduction}

Diffusive (Fermi) acceleration of charged particles in collisionless shocks
is believed to be the mechanism
responsible for the production of cosmic-rays, as well as for the non-thermal
emission from a wide variety of high energy astrophysical sources
\citep[For reviews see, e.g.,][]{Drury83,Blandford87,Axford94}.
A theory of collisionless shocks based on first principles is, however, lacking.
One of the major issues which are not understood is magnetic field amplification.
Post-shock (downstream) magnetic field strengths derived from recent observations
of supernova remnant shocks are significantly higher than those expected from shock
compression of the pre-shock (upstream) plasma, suggesting a significant
amplification of the pre-shock
magnetic field beyond compression \citep[e.g.][]{Voelk05}. This amplification
is most likely intimately related to the process of particle acceleration,
and is not understood theoretically.

GRB afterglow shocks present an extreme example of magnetic field amplification.
Phenomenological considerations suggest that the afterglow is produced by synchrotron
emission by electrons accelerated to high energy in a relativistic collsionless shock
driven by the GRB explosion into the medium surrounding the progenitor \citep[see,
e.g.][for reviews]{Zhang04,Piran05}. The post-shock magnetic field is inferred to be near
equipartition. If the shock propagates into an inter-stellar medium with characteristic
field amplitude of $\simeq1\,\mu$G, this implies amplification of the field energy
density (beyond compression) by $\sim7$ orders of magnitude
\citep{Gruzinov99,Gruzinov01}. While the amplification of the field to near equipartition
by electro-magnetic plasma instabilities appears likely
\citep[e.g.][]{Gruzinov99,Medvedev99}, such instabilities tend to create a field varying
on a plasma skin depth scale, $c/\omega_p$, which is expected to decay in the post-shock
plasma at a distance of a few skin depths away from the shock. Observations indicate,
however, that the field survives over a scale many orders of magnitude larger than
$c/\omega_p$. This suggests that the characteristic scale of field variation grows to
values much larger than $c/\omega_p$, and the challenge is, thus, to explain the
formation of equipartition field on a scale $\gg c/\omega_p$
\citep{Gruzinov99,Gruzinov01}. Various groups have recently attempted to address this
challenge using numerical plasma simulations
\citep{Silva03,Frederiksen04,Jaroschek04,Nishikawa05,Medvedev05,Kato05,Spitkovsy06}.
Since the calculations are extremely demanding numerically, the simulation boxes are
typically only a few tens of skin-depths wide and a clear picture of field length scale
growth has not yet emerged.

GRB afterglows provide a unique opportunity for diagnosing collisionless shock physics,
as they allow to observe a rapid evolution of the synchrotron spectrum over a wide span
of wavelengths and for a wide range of shock Lorentz factors.
Afterglow observations were used to constrain the downstream field, and the accelerated
electron energy distribution, both at high \citep{W97a,Freedman01} and low
\citep{W97b,EichlerWaxman} energy. Here we point out that afterglow
observations also provide constraints on the upstream field. A characterization of
the upstream field provides information not only on the circum burst medium,
but also on the collisionless shock physics. Amplification of
magnetic field fluctuations in the pre-shock plasma are naturally expected within
the frame work of diffusive (Fermi) shock acceleration of particles
\citep{Bell78,Blandford78}, as the high energy particles stream ahead of the shock.
Some evidence for such enhancement has been obtained from radio observations
of supernova remnants \citep[e.g.][]{Achterberg94}. Recently, it had been proposed
that the streaming of high energy particles in non-relativistic collisionless
shocks may significantly amplify not only the fluctuations in the magnetic field, but
also the (overall) amplitude of upstream field strength \citep{Bell04}. If such
amplification is indeed achieved, it would allow acceleration of particles to higher
energy. For supernovae, e.g., this may allow acceleration of particles beyond
$10^{15}$~eV, and for GRBs it may allow acceleration to ultra-high ($>10^{19}$~eV)
energy in the afterglow shock
\citep[and not only in the internal shocks, see][for review]{Wrev}.

We show in \S~\ref{sec:theory} below that X-ray afterglow observations
may be used to put constraints on the upstream field amplitude.
In \S~\ref{sec:application} we apply the results derived in \S~\ref{sec:theory}
to afterglow observations, and derive lower limits on the upstream
magnetic field for several GRBs. In \S~\ref{sec:discussion} we summarize
our results and discuss their implications, including the implications
for numerical simulations and for recent discussions of upstream
magnetic fields in relativistic GRB afterglow shocks \citep{Lyubarsky05,Nakar05}.

\section{Shock acceleration and maximum synchrotron energy}
\label{sec:theory}

Within the diffusive shock acceleration framework, high energy particles
cross the shock front multiple times, and gradually gain energy as in each
crossing (from up- to down-stream or vice versa) they are scattered by a
macroscopic flow {\it approaching} them. We first derive in \S~\ref{sec:acc}
a lower-limit to the acceleration time for a given up-stream magnetic field.
In \S~\ref{sec:loss} we estimate the energy loss time of high energy electrons
due to inverse-Compton scattering of afterglow photons. The maximum energy
to which electrons can be accelerated, and the maximum energy of emitted
synchrotron photons is then estimated, for a given upstream field, by comparing
the acceleration and loss times.

\subsection{Acceleration time}
\label{sec:acc}

Consider a relativistic shock expanding into the circum-burst medium, with
post-shock fluid of Lorentz factor $\Gamma\gg1$ relative to the upstream
medium. The high energy electrons that cross the shock from the downstream
region to the upstream region are confined (due to the Lorentz boost)
in the upstream fluid frame to a narrow cone around the shock normal of opening
angle $\sim1/\Gamma$. The residence time of such electrons in the upstream, i.e.
the time that the electrons spend in the upstream before being scattered back
into the downstream, is approximately given by the time it takes for the
electrons to be deflected by an angle $\sim1/\Gamma$. Once the high energy
electrons travel at an angle, with respect to the shock normal, which is
larger than $1/\Gamma$, they are overtaken by the shock. Thus, the upstream
residence time may be estimated as
\begin{equation}
\label{eq:t_u}
t_u\sim\gamma_e m_e c/\Gamma eB_{u}=\gamma_e' m_e c/eB_{u}.
\end{equation}
Here, $B_u$ is the characteristic upstream field amplitude, $\gamma_e$ is
the electron Lorentz factor measured in the upstream fluid frame, and
$\gamma_e'$ is the electron Lorentz factor measured in the downstream
fluid frame. From here on, primed variables would denote parameter values
measured in the downstream frame.

The acceleration time to energy $\gamma_e m_e c^2$ is given by the sum
of the residence times in the upstream and downstream over the many cycles
of upstream/downstream scattering the electron undergoes as its energy is
increased. Nevertheless, the acceleration time is not much larger than
the upstream residence time, $t_u$. This is due to two reasons. First, under the
assumption that the pre-shock magnetic field is not amplified, and its
amplitude is that present in the pre-shock plasma, the much stronger,
near equipartition magnetic field in the downstream deflects the electron
much faster than it is being deflected in the upstream. This implies that
a single "cycle time," i.e the time for back and forth crossing of the shock, is
dominated by the upstream residence time. Second, since the electron's energy
is increased by some (energy independent) factor in each cycle
\citep[e.g.][]{Gallant99}, the fact that
the upstream residence time increases linearly with the electron's energy implies
that the acceleration time is dominated by the last cycle time. Thus, the
acceleration time may be written as
\begin{equation}
\label{eq:tacc}
t_a=g\frac{\gamma_e'm_ec}{eB_{u}},
\end{equation}
where $g$ is a correction factor, with weak (logarithmic) dependence
on $\gamma_e$. The exact value of $g$ depends on the detailed
assumptions regarding the structure of the field and the scattering
process. For reasonable assumptions on the magnetic field, values as
small as $g\approx10$ may be obtained \citep[e.g.][]{Lemoine03,Lemoine06}.
We will conservatively adopt this value in what follows.

\subsection{Energy loss time and maximum synchrotron energy}
\label{sec:loss}

The maximum energy of accelerated electrons is limited by several factors.
First, it is limited by the time available for acceleration, which is
comparable to the shock expansion time. Balancing $t_a$ given by
eq.~(\ref{eq:tacc}) with the expansion time, $\Gamma^2t$ where $t$ is
the time measured by a distant observer, sets an upper limit to the electron
Lorentz factor,
\begin{equation}
\gamma_{\max}'\simeq \frac{eB_{u}\Gamma^2t}{gm_ec}
\sim10^7g_1^{-1}\Gamma_2^2t_3B_{\mu G}.
\end{equation}
Here $B_{u}=1B_{\mu G}~\mu$G, $g=10^1g_1$, $\Gamma=10^2\Gamma_2$,
and $t=10^3t_3$~s.

A more stringent constraint is obtained by considering the electrons'
energy loss. Under the assumption that the upstream magnetic field
is not significantly amplified, synchrotron cooling in the upstream is
negligible. Electrons in the upstream loose energy primarily by
inverse-Compton (IC) scattering of afterglow synchrotron photons
emitted in the downstream region. Since afterglow modelling typically
implies that the energy density in radiation is similar (or larger) than the
downstream magnetic field energy density, the large residence time of
electrons in the upstream implies that IC losses of electrons in the
upstream region is the dominant energy loss. This conclusion is valid,
of course, provided IC scattering is not deep in the Klein-Nishina (KN)
regime.

A comparison of the acceleration and energy loss time is most easily carried out in the
downstream frame. In this frame, the acceleration time is $t_a'\simeq t_a/\Gamma$. The
energy loss time may be estimated as follows. The energy density in synchrotron radiation
is similar in the upstream and in the downstream regions, and both can be denoted as
$U_{ph}'$. We show below that the KN effect is not important. Neglecting the KN effect, the
cooling time due to IC scattering of (downstream emitted) afterglow synchrotron photons
is
\begin{equation}
\label{eq:tc} t_{c}'\simeq\frac{3 m_ec}{4\sigma_TU_{ph}'\gamma_e'}.
\end{equation}
It is instructive to note that a similar estimate of $t_c$ may be obtained by considering
the energy loss rate in the upstream frame. In this frame, the angular distributions of
both photon and electron momenta are concentrated within a narrow cone of opening angle
$1/\Gamma$ around the shock normal, while the photon number density and (individual)
photon and electron energies are larger by a factor $\Gamma$ compared to their downstream
values, $n_{ph}\simeq\Gamma n'_{ph}$, $\nu\simeq\Gamma\nu'$,
$\gamma_e\simeq\Gamma\gamma_e'$. The energy loss rate in the upstream frame is
approximately given by $\dot{E}\sim n_{ph}\sigma_T
c(1-\cos1/\Gamma)\times\Gamma\gamma^{\prime2}_eh\nu'\sim n'_{ph}\sigma_T
c\gamma^{\prime2}_eh\nu'$, where the collision rate is corrected by a factor
$(1-\cos1/\Gamma)\approx1/2\Gamma^2$ since the characteristic angle between photon and
electron momenta is $1/\Gamma$, and $\Gamma\gamma_e^{\prime2}h\nu'$ is the characteristic
(upstream) energy of a scattered photon. Thus, the energy loss rates in both up- and
down-stream frames are similar, $\dot{E}\sim\dot{E}'\sim\gamma_e^{\prime2}\sigma_T
cU_{ph}'$, which implies $t_c\sim\Gamma t_c'\sim \Gamma m_e c/\gamma_e'\sigma_T U_{ph}'$
as obtained in eq.~(\ref{eq:tc}).

Comparing $t_c'$ and $t_a'\simeq t_a/\Gamma$ given by eq.~(\ref{eq:tacc})
we find
\begin{equation}\label{eq:maxLF}
\gamma_{\max}'\simeq\pfrac{3 e B_{u}\Gamma}{4g\sigma_TU_{ph}'}^{1/2}
\simeq8.5\times10^4\pfrac{ B_{\mu G}\Gamma_2}{g_1U_{ph,0}'}^{1/2},
\end{equation}
where $U_{ph}'=10^0U_{ph,0}'$~erg~cm$^{-3}$. Denoting by $f$ the fraction of shock
accelerated electron energy converted to synchrotron radiation, we have
$U_{ph}'=6f\eps_{e,-1}\Gamma_2^2n_0{\ \rm erg~cm^{-3}}$ where $n=10^0n_0{\ \rm cm^{-3}}$
is the proper density of the upstream plasma and $\eps_e=10^{-1}\eps_{e,-1}$ is the
fraction of downstream thermal energy carried by electrons. As we show below $f$ is  of
order unity and hence $U_{ph,0}'\sim1$.

In what follows we provide a detailed derivation of $f$, and hence of $U_{ph}'$. It is
important to note here that when the KN correction is taken into account, $U_{ph}'$
should be replaced by $U_{ph}'(<\nu_{KN})$, where $h\nu_{KN}=\Gamma mc^2/\gamma_{\max}'$
is the photon energy above which IC cooling becomes less efficient due to the KN effect.
For the derivation that follows, it is useful to define the downstream Compton $Y_{\max}$
parameter for electrons with Lorentz factor $\gamma_{\max}'$ as
$Y_{\max}=[U_{ph}'(<\nu_{KN})/U_B']_d$,  and replace $U_{ph}'$ in eq.~(\ref{eq:maxLF})
with $U_{ph}'=(U_B')_dY_{\max}$. The characteristic frequency of synchrotron photons
emitted by electrons with $\gamma_{\max}'$, $\nu_{\max}\simeq0.29\Gamma\gamma_{\max}'^2 e
B_d'/2\pi m_ec$, is then given by eq.~(\ref{eq:maxLF}) to be
\begin{eqnarray}\label{eq:maxnu}
\nu_{\max} &\simeq&2.9\times10^{18}\frac{ B_{\mu
G}}{g_1(\eps_{B,-2}n_0)^{1/2}}\frac{\Gamma_2}{Y_{\max}}{\rm Hz}\nonumber\\
&\simeq& 2.7\times10^{17}\frac{ B_{\mu
G}E_{53}^{1/8}}{g_1\eps_{B,-2}^{1/2}n_0^{5/8}}Y_{\max}^{-1}t_{d}^{-3/8}{\rm
Hz}.
\end{eqnarray}
Here $t=1t_d$~day, $\eps_B=10^{-2}\eps_{B,-2}$ is the fraction of downstream thermal
energy carried by the downstream magnetic field, and $E=10^{53}E_{53}$~erg is the
(isotropic equivalent) explosion energy. The second equality holds for the case of
uniform density circum burst medium, for which the Lorentz factor drops with radius $R$
following the Blandford-McKee self-similar solution, $\Gamma=(17E/16\pi
nm_pc^2)^{1/2}R^{-3/2}$ (Blandford \& McKee 1976), and the relation between $\Gamma$ and
the observer time $t$ is $t=R/4\Gamma^2c$ (Waxman 1997c). Note also that we have assumed
an isotropic electron distribution and fully tangled magnetic field downstream, and taken
into account the fact that the synchrotron radiation peaks at 0.29 times the gyration
frequency of the relevant electrons.

Let us finally derive the value of $Y_{\max}$ with consideration of the KN effect. In
order to estimate $Y_{\max}$ we first discuss the energy distribution of afterglow
photons, which is determined by the energy distribution of shock accelerated
electrons. Afterglow observations imply that the post-shock electron energy
distribution follows a power law, $dn_e/d\gamma_e\propto\gamma_e^{-p}$ for
$\gamma_{m}\leq\gamma_e\leq\gamma_{\max}$, with a spectral index $p\approx2.2$ (Waxman
1997a, Freedman \& Waxman 2001, and references therein; Wu \etal 2004). This energy
distribution is consistent with the theoretical value derived for isotropic diffusion
of accelerated particles (in the test particle limit) in both numerical calculations
(Bednarz \& Ostrowski 1998; Kirk \etal 2000; Achterberg \etal 2001) and analytic
analyses (Keshet \& Waxman 2005). The synchrotron spectrum of such an electron
distribution follows a power-law, with power per unit frequency
$f_\nu\propto\nu^{-(p-1)/2}$, up to the energy where the radiative (synchrotron and
IC) losses of the electrons become important. This occurs at electron energy for which
the radiative loss time is shorter than the adiabatic cooling time, i.e. than the
energy loss time due to the expansion of the post shock plasma. The adiabatic loss
time, $t_{ad}'\simeq 6R/13c\Gamma$ (Gruzinov \& Waxman 1999) with $R=4\Gamma^2ct$ (for
uniform circum burst density), is longer than the radiative cooling time for electrons
with Lorentz factors exceeding $\gamma_c'=3mc/4\sigma_T(U_B')_d(1+Y_c)t_{ad}'$. Here,
$Y_c$ is the Compton $Y$ parameter for electrons with $\gamma_e'=\gamma_c'$. The
characteristic synchrotron frequency of photons emitted by electrons with
$\gamma_e'=\gamma_c'$, $\nu_c\simeq0.29\Gamma\gamma_c'^2 e B_d'/2\pi m_ec$, is (for
uniform circum burst density)
\begin{eqnarray}
\nu_c&\simeq&\frac{2.5\times10^{13}}{(\eps_{B,-2}n_0)^{3/2}(1+Y_c)^2\Gamma_2^4t_3^2}{\rm
Hz}\nonumber\\
&\simeq&\frac{7.5\times10^{13}}{\eps_{B,-2}^{3/2}E_{53}^{1/2}n_0}
\frac {t_{d}^{-1/2}}{(1+Y_c)^{2}}{\rm Hz}.
\end{eqnarray}
At higher frequency, the photon spectrum steepens from
$f_\nu\propto\nu^{-(p-1)/2}\approx\nu^{-1/2}$
to $f_\nu\propto\nu^{-p/2}\approx\nu^{-1}$.

In order to complete the description of the synchrotron photon spectrum, let us estimate
the lowest energy of accelerated electrons. The fraction of downstream thermal energy
carried by electrons, $\eps_e$, is typically $\eps_e\gtrsim0.1$ (Freedman \& Waxman
2001). This implies a minimum Lorentz factor of accelerated electrons of
$\gamma_m'\simeq\eps_e(m_p/m_e)\Gamma$, for which the characteristic frequency of
synchrotron emission is
$\nu_m\simeq10^{13}\eps_{e,-1}^2(\eps_{B,-2}n_0)^{1/2}\Gamma_1^4$~Hz. For $t>1$~hr the
afterglow is typically in  the "slow cooling regime," with $\gamma_m'<\gamma_c'$. The
synchrotron spectrum is therefore well approximated by a broken power law, with
$f_\nu\propto\nu^{-(p-1)/2}\approx\nu^{-1/2}$ for $\nu_m<\nu<\nu_c$ and
$f_\nu\propto\nu^{-p/2}\approx\nu^{-1}$ for $\nu>\nu_c$. In this case, the energy density
of synchrotron photons is mainly concentrated around $\nu_c$, i.e. $U_{syn}\simeq
U_{ph}(<\nu_c)$, and we may approximate
\begin{equation}
Y_{\max}=Y_{syn}\times\min[1,~(\nu_{KN}/\nu_c)^{1/2}],
\end{equation}
where $Y_{syn}\equiv (U_{syn}'/U_B')_d$ is the downstream energy
density ratio between synchrotron radiation and magnetic field and
$\nu_{KN}$ is the frequency of photons for which IC scattering of
electrons with $\gamma_e'=\gamma_{\max}'$ is in the
the KN regime. $\nu_{KN}=\Gamma m_ec^2/h\gamma_{\max}'$, i.e.
\begin{eqnarray}
\nu_{KN}&=&1.3\times10^{17}\pfrac{g_1\eps_{B,-2}n_0}{B_{\mu
G}}^{1/2}\Gamma_2^{3/2}Y_{\max}^{1/2}{\rm Hz}\nonumber\\
&\simeq&3.1\times10^{15}\pfrac{g_1\eps_{B,-2}}{B_{\mu
G}}^{1/2}E_{53}^{3/16}n_0^{5/16} t_{d}^{-9/16}Y_{\max}^{1/2}{\rm
Hz}.
\end{eqnarray}
Here too, the second equality holds for uniform density circum burst medium.

Using the equations for $\nu_{KN}$ and $\nu_c$ we have
\begin{equation}
\frac{\nu_{KN}}{\nu_c}=41\frac{g_1^{1/2}\eps_{B,-2}^2E_{53}^{11/16}n_0^{21/16}}{B_{\mu
G}}\frac{(1+Y_c)^2Y_{\max}^{1/2}}{t_{d}^{1/16}}.
\end{equation}
This implies that unless the upstream field is amplified to $B_{\mu G}\gg1$,
$\nu_c\lesssim\nu_{KN}$. In this case we may approximate $Y_{\max}\approx Y_{syn}$. As for
$Y_{syn}$, if $Y_{syn}\ga1$ and only single IC scattering is considered (multiple IC
scattering is typically suppressed as it is well within the KN regime), then
$Y_{syn}\simeq(\eta\eps_e/\eps_B)^{1/2}$, where $\eta=(\nu_c/\nu_m)^{-(p-2)/2}$ is the
"radiative efficiency" of the post-shock electrons (e.g. Sari \& Esin 2001). For
$p\approx2$ we have $\eta\approx1$, and since afterglow observations imply $\eps_e\ga0.1$,
we find $\eta\eps_e/\eps_B\ga1$ and $Y_{syn}\ga1$. To summarize, afterglow observations
imply $Y_{\max}\approx
Y_{syn}\approx(\eta\eps_e/\eps_B)^{1/2} \approx(\eps_e/\eps_B)^{1/2}\sim$~a few.

With $Y_{\max}\approx(\eta\eps_e/\eps_B)^{1/2}$, eq.~(\ref{eq:maxnu}) then implies a
maximum afterglow synchrotron photon energy of
\begin{equation}
\label{eq:maxphoton}
h\nu_{\max}^{\rm ob.}\simeq0.3 \frac{ B_{\mu
G}E_{53}^{1/8}}{g_1(\eta\eps_{e,-1})^{1/2}[(1+z)n_0]^{5/8}}t_{d}^{-3/8}{\rm keV },
\end{equation}
where we have taken here into account the redshift $z$ of the source.
Note, that $\nu_{\max}^{\rm ob.}$ is independent of the poorly known $\eps_B$.
At photons energies larger than $h\nu_{\max}^{\rm ob.}$, an exponential
cut-off is expected in the afterglow spectrum. Since for $B_{\mu G}\sim1$
this cut-off is expected to take place in the soft X-ray band, X-ray afterglow
observations provide an interesting constraint on the upstream field.
If the synchrotron spectrum is
observed to extend beyond an energy $h\nu^{\rm ob.}$, then a lower limit to
the upstream field is implied,
\begin{equation}\label{eq:lowerB}
B_{u}>50 \frac{g_1(\eta\eps_{e,-1})^{1/2}}{E_{53}^{1/8}}\left
(n_0\frac{1+z}2\right)^{5/8}t_{d}^{3/8}\frac{h\nu^{\rm ob.}}{10\,\rm keV}
\mu{\rm G }.
\end{equation}

\section{Lower limits on upstream field strength}
\label{sec:application}

The BeppoSAX catalog of X-ray afterglows is presented in De Pasquale \etal (2005).
X-ray data are typically available between $\sim0.3$~d and $\sim1$~d. The
0.1 to 10~keV spectra are well fit by a simple photoelectrically absorbed
power-law. As shown in their Fig. 1 of De Pasquale \etal (2005), most of the spectra
have been well determined up to 10~keV, with no sign of a cut-off or steepening of
the spectra at 10~keV. This implies, using eq.~(\ref{eq:lowerB}),
\begin{equation}\label{eq:BeppoSax}
B_{u}\gg50 \frac{g_1(\eta\eps_{e,-1})^{1/2}}{E_{53}^{1/8}}\left
(n_0\frac{1+z}2\right)^{5/8} \mu{\rm G }.
\end{equation}
This lower limit is only weakly dependent on $E$, but
somewhat sensitive to $\eps_e$. Values of $\eps_e\gtrsim0.1$ are inferred
from afterglow modelling, and are consistent with the clustering of
explosion energy (Frail \etal 2001)
and X-ray afterglow luminosity (Freedman \& Waxman 2001).

In several cases, stronger constraints on the magnetic field are obtained.
\\
{\it GRB 990123}. This is the only case where the afterglow was detected up to photon
energy of 60 keV (using the BeppoSAX PDS instrument, Maiorano \etal 2005; Corsi et al.
2005), implying non-thermal emission up to this high energy at $\simeq.5$~d, and a limit
on the magnetic field of $B_u>0.2n_0^{5/8}$~mG.
\\
{\it GRB 030329.}
In this case, there are very late-time detections of X-ray
emission at 37, 61 and 258 days by XMM-Newton (Tiengo et al.
2003). Since at late-times the expansion may not be highly
relativistic, as assumed in the analysis of this paper, and since
late time emission could be affected by the associated
supernova, we only consider here the observation at 37 days,
with $h\nu_{\max}^{\rm ob.}>10$~keV. Eq.~(\ref{eq:lowerB}) then
implies $B_u>0.2n_0^{5/8}$~mG. We note that this numerical value is
an underestimate, since at this late time the shock expansion is
not spherical (due to jet expansion), leading to a faster decrease
of $\Gamma$ than assumed in our analysis.
\\
{\it GRB 050904.}
For this high-$z$ burst ($z=6.3$), the Swift XRT (0.2-10 keV)
observation at $3\times10^5$~s shows emission up to hard X-rays,
1.4-73 keV in the source rest frame (Cusumano
\etal 2005). Eq.~(\ref{eq:lowerB}) then implies
$B_u>3(n/100{\rm cm^{-3}})^{5/8}$~mG~$=0.2n_0^{5/8}$~mG.
Here we have given a value also for large $n$, as the analysis of
absorption lines in the optical afterglow implies
$n_e=10^{2.3\pm0.7}{\rm cm}^{-3}$ (Kawai \etal 2005).

\section{Discussion}
\label{sec:discussion}

We have shown that late time, $t\gtrsim1$~d, X-ray observations of GRB afterglows
provide interesting constraints on the upstream magnetic field. A lower limit to
the magnetic field is obtained by requiring the acceleration time of electrons
producing high energy synchrotron photons to be shorter than their energy loss
time due to inverse-Compton scattering of afterglow photons. The lower limit for
the magnetic field is given in eq.~(\ref{eq:lowerB}) as a function of the energy of the
observed high energy photons, $h\nu$. We have then shown in \S~\ref{sec:application}
that X-ray afterglow observations typically imply an upstream field strength of
$B\gg0.05n_0^{5/8}$~mG, where $n=10^0n_0{\rm cm^{-3}}$ is the upstream plasma density.
In several cases, $B>0.2n_0^{5/8}$~mG is obtained.

The large magnetic fields inferred are unlikely to be present in the interstellar medium
(ISM) of the galaxies hosting the GRBs. This is easy to see by assuming equipartition
between the magnetic field energy density and the turbulent energy density in the ISM,
$B^2/8\pi=n m_p v_T^2/2$, which implies that the turbulent velocity required to support
the inferred $B>0.2n_0^{5/8}$~mG field is $v_T\gtrsim500 n_0^{1/8}{\rm km/s}$. Indeed,
mG magnetic fields are observed in star burst galaxies \citep{Thompson05}, but the ISM
density in such galaxies is high, $n_0\gg1$, implying that the constraint
$B>0.2n_0^{5/8}$~mG is not satisfied.

Given the above estimate of $v_T\gtrsim500 n_0^{1/8}{\rm km/s}$, it is clear that a
sufficiently strong magnetic field may be present in a wind emitted by the GRB
progenitor, and into which the shock propagates, provided that the wind is fast,
$v_w\gtrsim10^3{\rm km/s}$, and highly magnetized, i.e. provided that it carries magnetic
field energy density which is not much smaller than its kinetic energy density. Such
interpretation faces two challenges. Wolf-Rayet stars posses fast, $\sim10^3{\rm km/s}$,
winds, and are the likely progenitors of SN of type Ic, which are associated with GRBs
\citep[e.g.][]{Zhang04,Piran05}. However, such hot stars have radiative envelopes and are
not expected therefore to have magnetically driven winds. Nevertheless, the possibility
that magnetic fields play an important role in the winds of such hot stars can not be
ruled out \citep[e.g.][]{Poe89,Ignace03}. The second challenge is that afterglow
observations have not provided so far a clear indication for GRB shocks propagating into
a circum burst wind \citep[e.g.][]{Zhang04}. However, it should be noted that late time,
$\gtrsim1$~d, afterglow observations do not allow to clearly distinguish between
propagation into a wind and propagation into uniform density characteristic of the ISM
\citep[e.g.][]{Livio00}, and that early time afterglow observations provided by Swift are
not yet properly understood \citep[e.g.][]{Fan06}.

If the GRB shock is not propagating into a highly magnetized wind, then the large
upstream magnetic fields inferred by our analysis require strong amplification
of the field ahead of the shock, most likely by the streaming of high energy
particles. If this is the correct interpretation, then numerical simulations of
relativistic, weakly magnetized collisionless shocks that properly describe the
shock structure should present strong pre-shock magnetic field amplification,
and thus also particle acceleration (note, that the precursor magnetic field
obtained in the simulation of Spitkovsy 2006 is a numerical effect, due to
reflection of particles from the edge of the simulation box). Finally, it has
recently been concluded by \citet{Lyubarsky05} that the saturation of the Weibel
instability implies that the shock is mediated by deflections of particles in the
pre-existing, $\sim1\mu{\rm G}$ upstream field, and Milosavljevi\'c \& Nakar (2005)
have postulated in their phenomenological analysis interaction of the accelerated
particles with the pre-existing $\sim1\mu{\rm G}$ upstream field. The constraints
derived by our analysis indicated that this conclusion (postulation) may not apply
to collisionless GRB shocks.

\acknowledgments
This work is partially supported by an ISF grant.

\end{document}